\documentclass[aps,prb,twocolumn,groupedaddress]{revtex4-1}
\usepackage{graphicx}  
\usepackage{amssymb}  
\usepackage{array}
\usepackage{amsmath}
\usepackage{textcase}
\usepackage{bm}
\usepackage{csquotes}
\usepackage{braket}
\usepackage[colorlinks=true,linkcolor=blue,citecolor=blue,urlcolor=blue,breaklinks]{hyperref}
\usepackage[subrefformat=parens,labelformat=parens, caption=false]{subfig}

\begin{document}
\title{Probing the Origins of Inhomogeneous Broadening
	in Nitrogen-Vacancy Centers with \enquote{Doppler-Free} Type Spectroscopy}

\author{Y. Rosenzweig$^{1}$}
\email {yosefros@post.bgu.ac.il}
\author{Y. Schlussel$^{1}$}
\author{R. Folman$^1$}
\affiliation{$^1$ Department of Physics, Ben-Gurion University of the Negev, Be'er Sheva 84105, Israel}
\date{\today}

\begin{abstract}  
In order to better understand the underlying fundamental physical processes in nitrogen-vacancy centers in diamond, and to realize improvements in the use of this system for technological applications, it is imperative to gain new insight into the origins of the apparent inhomogeneous broadening. In this work we make use of a novel type of spectroscopy developed specifically for this task. The pump-probe spectroscopy closely follows Doppler-free spectroscopy used in atomic vapor. We show that the origin of inhomogeneous broadening comes from local magnetic field variations in the diamond lattice. 
\end{abstract}

\pacs{}
\maketitle

\section{Introduction}
\label{Introduction}

The negatively charged  nitrogen-vacancy center in diamond (NV) is formed by a nitrogen atom adjacent to a vacant site in a carbon diamond lattice. Although the NV is within a solid-state crystal, it has \enquote{atom-like} properties such as discrete energy levels, spin state, and long coherence times \cite{DOHERTY2013,Awschalom1174}. The combination of solid-state and atom-like properties yields a simple and effective system for fundamental physics studies \cite{Hensen2015} as well as technological applications (e.g. sensing electric and magnetic fields) \cite{Wolf2015,Dolde2011}.
\par
When measuring transitions with an ensemble of NVs, the varying local environments of the NVs, due to strain as well as nearby ${}^{13}\textrm{C}$, substitutional nitrogen (P1) and other impurities, generate a broadened spectroscopic linewidth  \cite{bargil2012,Takahashi2008,Jensen2013}. For fundamental studies of NVs as well as the development of technological applications, it is of significant interest to better understand the main origin of this broadening.
\par
 In this work we improve the work presented in Ref.\,11, by utilizing two new techniques. First, we confine our experiment to a single hyperfine transition, compared to the wide scan in Ref.\,11, in order to avoid masking effects. Even though these results constitute an improvement over the results of Ref. 11, it is shown that they are still not sound enough due to inherent experimental fluctuations (Sec.\,\ref{saturation spectroscopy}). Thus, we introduce a novel spectroscopy technique (Sec.\,\ref{sec:doppler}), inspired by Doppler free spectroscopy in vapor, which is found to be much more immune to experimental fluctuations, enabling a more sound interpretation of the results. These techniques allow us to state with a high level of confidence that the main contributor to the transition broadening is the effect of local magnetic fields.

\section{NV Hamiltonian}
\label{NV Hamiltonian}
\begin{figure}
	\includegraphics[width=1\linewidth]{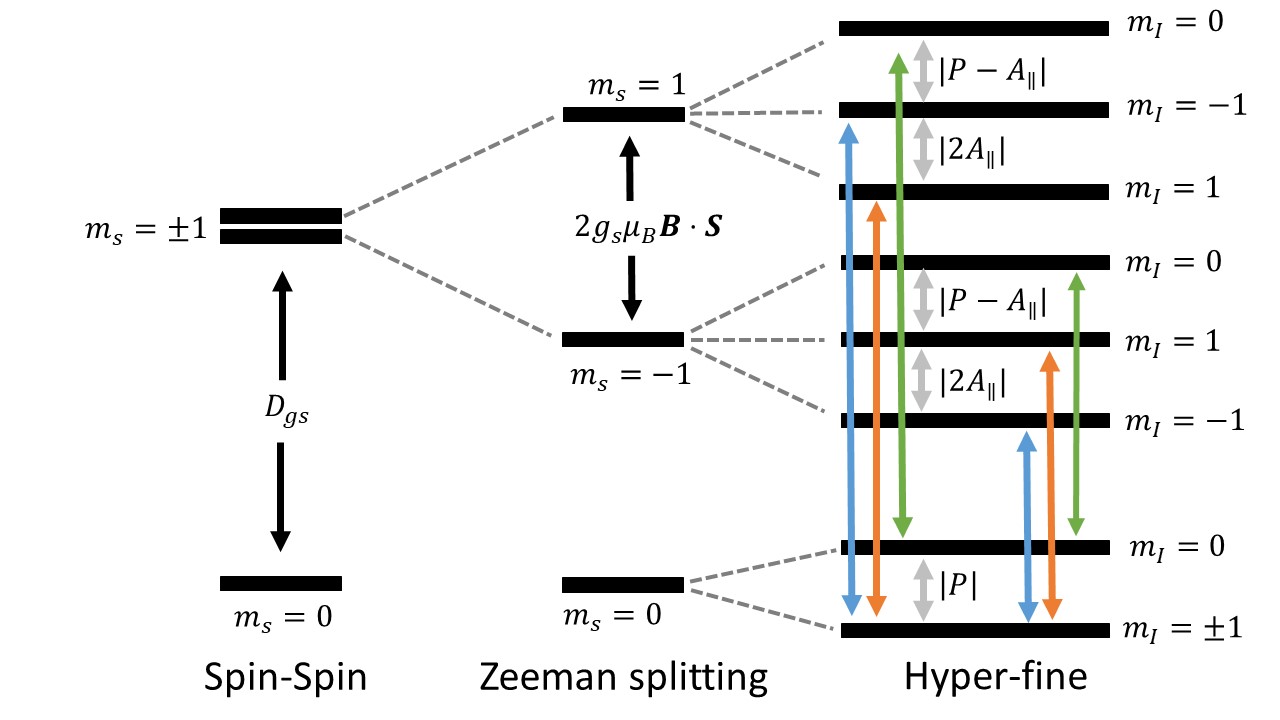}
	\caption{The energy levels of the NV ground state. The MW transitions from $m_s=0$ to $m_s=\pm1$ conserve the nuclear spin projection, $m_I$, as depicted by the colored arrows (green, blue and orange for $m_I=0$, $-1$ and $+1$, respectively). The gray arrows represent the energy spacing between the hyperfine states.}
	\label{fig:Energylevels}
\end{figure}

\begin{figure}
	\centering
	\includegraphics[width=1\linewidth]{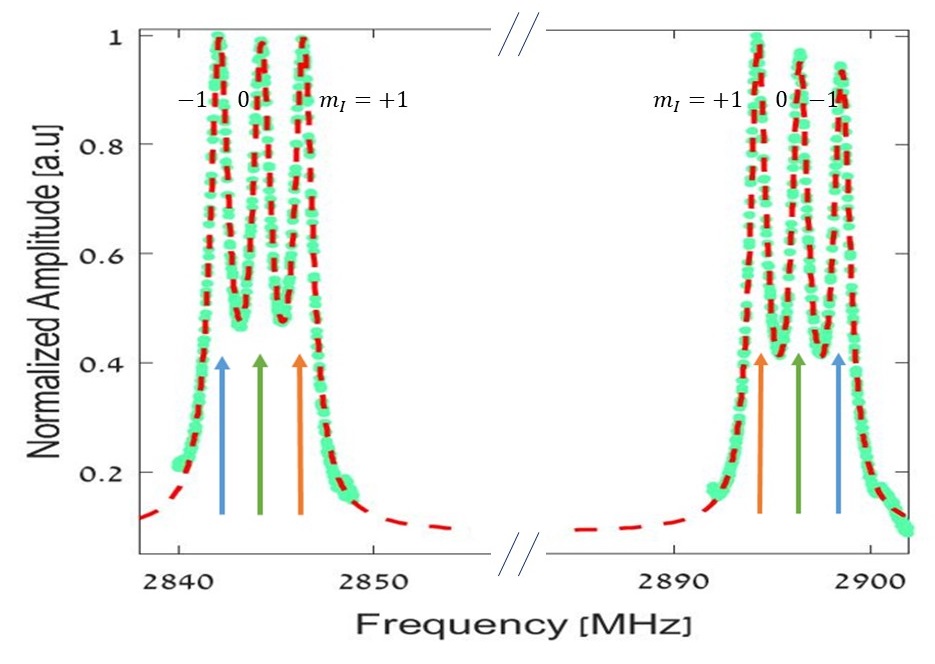}
	\caption{ODMR of a single NV orientation [111] using a lock-in amplifier with a static magnetic field along [111]. Data in turquoise and fit in red. The other 3 orientations of the NV are not shown. Arrows with the same color show a transition from $m_s=0$ to $m_s=\pm 1$ with the same $m_I$ number. The color code is the same as in Fig.~\ref{fig:Energylevels}.}
	\label{fig:ODMRbothsideswitharrowsV2}
\end{figure}

In order to understand the origin of line broadening in an ensemble of NVs we examine the Hamiltonian of a single NV. We neglect the off-diagonal zero field splitting $E(\textbf{S}_y^2-\textbf{S}_x^2)$. The latter can induce mixing of the Zeeman states at low external magnetic field. Exposed to an external magnetic field of several Gauss and above, the eigen states become \enquote{Zeeman states} as the relative mixing effect diminishes. The Hamiltonian then simply reads \cite{DOHERTY2013}
\begin{eqnarray}
\begin{aligned}
\mathcal{H}={} & 	D_\mathrm{gs}\textbf{S}_z^2+\gamma\textbf{B}\cdot\textbf{S}+P\textbf{I}_z^2+A_\parallel\textbf{S}_z\textbf{I}_z+
\\
& A_\perp(\textbf{S}_x\textbf{I}_x+\textbf{S}_y\textbf{I}_y)-g_N\mu _N\textbf{B}\cdot\textbf{I}\,	\label{eq:hamiltonian}
\end{aligned}
\end{eqnarray}
where $z$ is along the nitrogen-vacancy axis, $D_\mathrm{gs}=2.87\,\mathrm{GHz}$, $\gamma=2.8\,\mathrm{MHz/G}$, $P=-4.95\,\mathrm{MHz}$, $A_\perp\approx2.7\,\mathrm{MHz}$, $A_\parallel\approx2.16\,\mathrm{MHz}$ and $g_n\mu_N=0.31\mathrm{kHz/G}$ \cite{DOHERTY2013,Bermudez2011}. The electronic spin in NVs is $S=1$ and also the nuclear spin is $I=1$ as we take into account only the nucleus of $^{14}\textrm{N}$ (the natural abundance of $^{14}\textrm{N}$, which has a spin 1, is $99.6\,\%$). The three $m_I$ hyperfine projections cause a 3-fold splitting of each transition frequency between the Zeeman sub-levels. The transition frequencies of the 3 hyperfine projections within the $m_s=0$ to $m_s=\pm1$ transitions are symmetric to reflection with respect to $D_\mathrm{gs}$ so that the $m_I=-1$ transitions are the farthest from $D_\mathrm{gs}$ while the $m_I=+1$ transitions are the closest to it (as depicted in Fig.\,\ref{fig:Energylevels} and Fig.\,\ref{fig:ODMRbothsideswitharrowsV2}). A typical FWHM of the spectroscopic linewidth of these hyperfine transitions is $\sim1.1\,\mathrm{MHz}$ for our diamond at room temperature, much larger than a single NV linewidth \cite{Mizuochi2009}, which implies the existence of an inhomogeneous broadening mechanism. From the Hamiltonian it is evident that different NVs with different local environments (in terms of electric and magnetic fields) yield different resonance frequencies. This could clearly cause inhomogeneous spectroscopic line broadening \cite{pauli2014}.

\par
The local fields can be modeled \cite{pauli2014} by adding to the Hamiltonian in Eq.\,\eqref{eq:hamiltonian} a local magnetic field $\delta B$, a local electric (strain) field $\delta\epsilon$ and neglecting the nuclear spin. The latter simplification may be justified by the fact that the additional local fields manifest themselves mostly when interacting with the electronic spin [see Eq.\,\eqref{eq:hamiltonian}], causing the entire 3-fold hyperfine splitting of the Zeeman states to move as one. As the conclusions arrived at in later stages of this work deal with relative frequencies, the effective Hamiltonian neglecting the nuclear spin is valid. This Hamiltonian for axial fields now reads \cite{pauli2014}
\begin{eqnarray}
\mathcal{H}=(D_\mathrm{gs}+d^\parallel\delta\epsilon)S_z^2+\gamma(B_z+\delta B_z)S_z\,,
\label{eq:readuced Hamiltonian}
\end{eqnarray}
and the resonant transition frequencies from $m_s=0$ to $m_s=\pm1$ for each individual NV are
\begin{eqnarray}
f^0_{\pm}=(D_\mathrm{gs}+d^\parallel\delta\epsilon)\pm \gamma(B_z+\delta B_z)\,,
\end{eqnarray}
where $\gamma$ is the gyromagnetic ratio and $d^\parallel$ is the electric dipole along the z direction.
Eq.\,\eqref{eq:readuced Hamiltonian} is true for all 3 $m_I$ levels. Note that as shown in Fig.\,\ref{fig:ODMRbothsideswitharrowsV2}, each NV has three nearby values for $f^0_-$ as well as for $f^0_+$, due to the 3 values of $m_I$. Thus, the transition frequencies can be written as $f^0_{\pm_i}$  where $m_I=i$. 
\par
Let us now examine the relation between $f^0_{+_i}$ and $f^0_{-_i}$ under two extreme hypotheses: first, that the magnetic field is the main contributer to the inhomogeneous broadening and we can neglect the effect of the local electric field, or second, that the local electric field is the main contributer and we can neglect the local magnetic field. Under the first hypothesis, we neglect $d^\parallel\delta\epsilon$, and obtain the following relation
\begin{eqnarray}\label{eq:probevspump}
f^0_{+_i}=2D_\mathrm{gs}-f^0_{-_i}\,.
\end{eqnarray}
While this equality is hypothesized to be valid for all NVs, it should be again noted that for each NV, $f^0_{\pm_i}$ are different. Alternatively, under the second hypothesis if the local electric field is the main contributer, we can neglect $\delta B_z$ and the following relation would be valid:
\begin{eqnarray}\label{eq:probewithpump}
f^0_{+_i}=2\gamma B_z+f^0_{-_i}\,.
\end{eqnarray} 
Eqs.\,\eqref{eq:probevspump} and \,\eqref{eq:probewithpump} show how the location of $f^0_{-_i}$ is being dictated by the location of $f^0_{+_i}$. Let us define the mean ensemble transition frequency (i.e. center of the distribution) from $m_s=0$ to $m_s=\pm1$ with $m_I=i$ as $f^c_{+_i}$ and $f^c_{-_i}$. If we only slightly change $f^0_{+_i}$ from $f^c_{+_i}$ (namely, interact with NVs that have different local environment and hence a new $f^0_{+_i}$) so that we stay well inside the hyperfine transition linewidth, i.e., $f^0_{+_i}=f^c_{+_i} + \delta f$ where $\delta f<1\,\mathrm{MHz}$, there should also be a corresponding linear change in the conjugate frequency of the same NVs, such that if the magnetic contribution is dominant [Eq.\,\eqref{eq:probevspump}] $f^0_{-_i}$=$f^c_{-_i}-\delta f$, and if the electric contribution is dominant [Eq.\,\eqref{eq:probewithpump}] $f^0_{-_i}$=$f^c_{-_i}+\delta f$ [this difference is due to the fact that $\delta \epsilon$ is related to the $S_z^2$ terms in Eq.\,\eqref{eq:readuced Hamiltonian}, while $\delta B_z$ is related to $S_z$]. Consequently, as shown in the following, when $f^0_{+_i}$ moves to the right, $f^0_{-_i}$ moves to the left in the case of a dominant magnetic contribution, while it moves to the right in the case of a dominant electric contribution. This orthogonal behavior allows us in the following to discriminate between the two models. 

\section{setup}
\label{setup}
We first start by preparing all NVs in the $m_s=0$ state. When illuminating an $m_s=0$ NV center with green laser light, the NV is excited to a vibrionic level in the optically excited state, following which it decays rapidly to the lowest vibrionic level,while emitting a red photon. In contrast, when the system is optically excited from $m_s=\pm1$, there is a possibility for non-radiative decay through an intersystem crossing, which results in reduced fluorescence compared to the previous transition \cite{Manson2006,Goldman2015}. 
Thus, while the first transition conserves spin projection, the second does not. This enables optical pumping into the $m_s=0$ state. We may also use the above characteristics to optically distinguish between state $m_s=0$ and states $m_s=\pm1$. Specifically, applying a micro-wave (MW) field, on resonance with the ground state level splitting, will result in a drop in fluorescence, allowing us to optically detect the transition \cite{Robledo2011}.
\par
\begin{figure}
	\centering
	\includegraphics[width=1\linewidth]{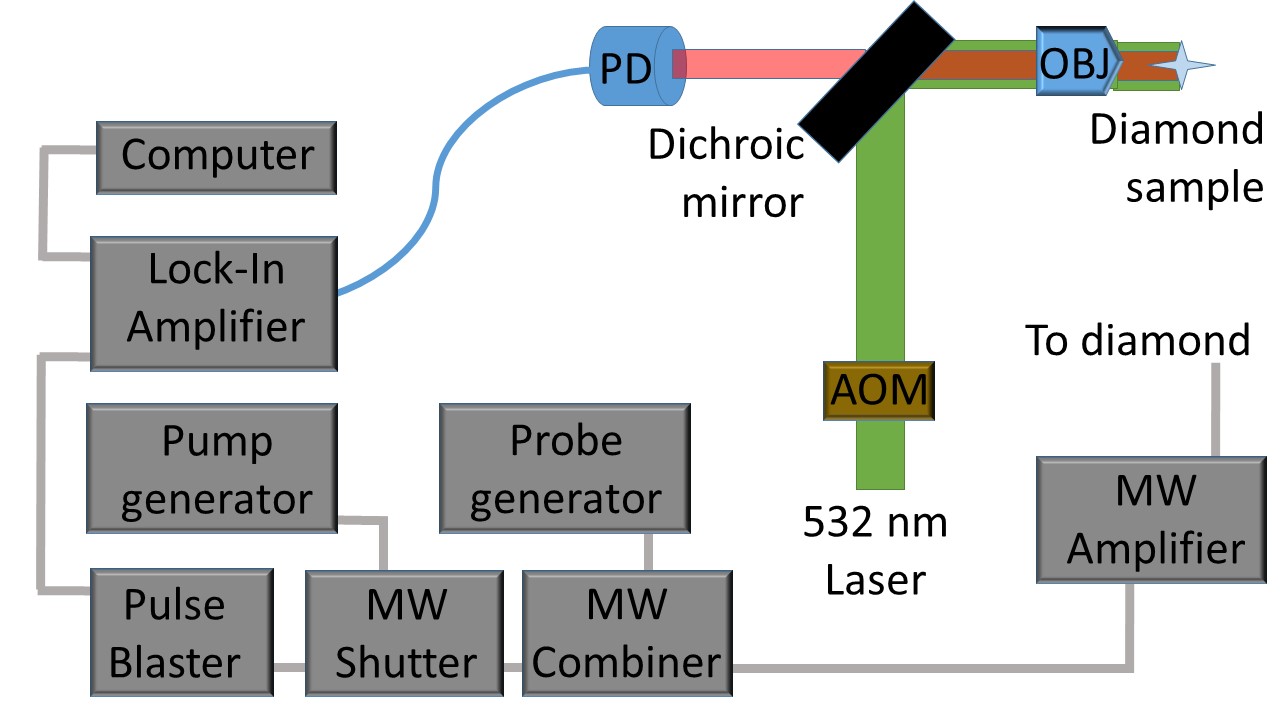}
	\caption[Experimental Setup]{Schematic diagram of the experimental setup. The laser beam is reflected from a dichroic mirror and focused by an objective lens (OBJ) onto the diamond, while the fluorescence traverses
	the OBJ and the dichroic mirror and is collected by the photo-diode (PD) with its output connected to a lock-in amplifier. Left: the MW pump generator is connected to a MW shutter which modulates the signal, and then the pump signal is combined with the MW probe signal. The combined signal is amplified and the total amplified signal (modulated pump and CW probe) is injected into a MW antenna near the diamond. The modulation rate of the MW shutter is controlled by a pulse generator that also gives the reference for the lock-in amplifier. All instruments are connected to the same computer.}
	\label{fig:setup}
\end{figure}

In the experiment (Fig.\,\ref{fig:setup}), we use a $532\,\mathrm{nm}$ green laser ($110\,\mathrm{mW}$ output). As is standard in NV experiments, the beam is reflected by a dichroic mirror and focused by an objective lens (Olympus, Pro-Plan $40\times$ magnification; N.A.=0.6) onto the diamond. The emitted red fluorescence collected by the objective is transmitted through the dichroic mirror to a photo-diode (ThorLabs, APD110AD), which is connected to a lock-in amplifier (SRS, SR844). The $m_s=0$ to $m_S=+1$ MW (R\&D SMR20, set to $-12\,\mathrm{dbm}$ output), referred to in the following as the \enquote{pump}, is modulated by a MW shutter (Mini-Circuits, ZASWA-2-50DR+) and then combined with the $m_s=0$ to $m_s=-1$ MW (SRS SG384, set to $-20\,\mathrm{dbm}$ output) referred to in the following as the \enquote{probe}, using a beam combiner (Mini-Circuits, ZB4PD-42), and finally both are amplified (Mini-circuits ZHL16W-43+). The total amplified signal (modulated pump and CW probe) is injected into a MW antenna near the diamond. The modulation rate of the MW shutter, $36.6\,\mathrm{kHz}$, controlled by a pulse generator (Pulse Blaster ESR-PRO $500\,\mathrm{MHz}$) is fed into the lock-in amplifier. Throughout the entire measurement, the laser beam is working in continuous mode (CW). The diamond which we use is a HTHP (High-Temperature-High-Pressure) type diamond, with an NV density of $10\,\mathrm{ppm}$.

\section{saturation spectroscopy}
\label{saturation spectroscopy}

To investigate whether Eq.\,\eqref{eq:probevspump}, Eq.\,\eqref{eq:probewithpump}, or some combination of the above dominates the behavior of the NV ensemble, we begin with high-resolution saturation spectroscopy in which we use an amplitude modulated MW pump field with a fixed frequency of $f_+$, and scan with a MW CW probe field of frequency $f_-$ across the 3 hyperfine states in the $m_s=-1$ transition (for same level saturation spectroscopy see for example Ref.\,15\nocite{Gawlik2016}). Notice that we assign an $i$ index for $f^0_{\pm}$ as it represents a transition frequency to a specific hyperfine transition, while $f_+$ ($f_-$) is simply the pump (probe) frequency. During the scan we measure the modulated fluorescence emitted from the diamond using a lock-in amplifier.
\par
\begin{figure}
	\centering
	\includegraphics[width=1\linewidth]{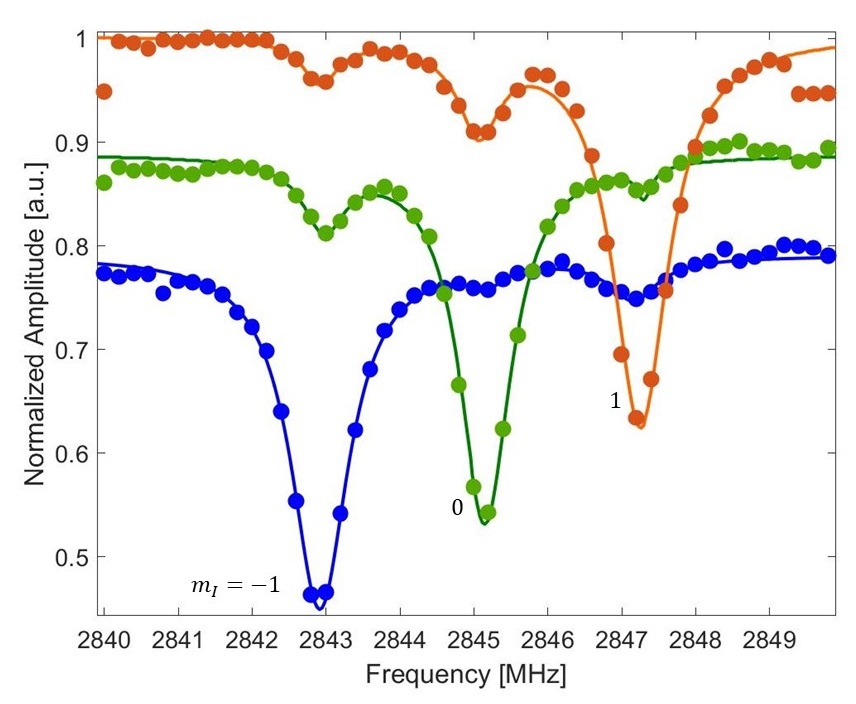}
	\caption[Spectroscopic Holes]{Spectroscopic saturation holes using a lock-in amplifier with a modulation frequency of $36.6\,\mathrm{kHz}$. Data in dots, fit in solid line. The color code is the same as in Figs.~\ref{fig:Energylevels} and \ref{fig:ODMRbothsideswitharrowsV2}: in the orange, green and blue the pump is set to the $m_I=+1,0,-1$ transitions, respectively. Data of $m_I=0,-1$ were shifted down after normalization for better clarity. The average linewidth of the saturation hole is $0.9\,\mathrm{MHz}$ which is $\sim200\,\mathrm{kHz}$ less than an ODMR width. This could be explained by the removal of broadening generated by slowly varying local magnetic fields (most likely $^{13}\textrm{C}$ impurities)\cite{pauli2014}. Similar results were obtained while modulating at $426\,\mathrm{Hz}$. Finally, while the observed side-dips could be due to some unknown physics (such as $m_I$ mixing transitions), several studies we made favor the option that they are due to MW sidebands produced by the MW amplifier.}
	\label{fig:3holewithfit}
\end{figure}

 If the pump field frequency is on resonance ($f_+=f^0_{+_i}$) and the probe field is far from resonance ($f_-\neq f^0_{-_i}$), we measure a high lock-in signal due solely to the modulated pump field: when the MW is off, $m_s=0$ is populated due to the green laser excitation, and when the MW field is on, state $m_s=+1$ is populated and the fluorescence is reduced. Thus, the resonance pump MW modulation generates fluorescence modulation with the same frequency, and an amplitude which is proportional to the fluorescence difference between the two ground states. As we bring the probe field closer to resonance ($f_-=f^0_{-_i}$), there are less NV centers available for excitation by the pump, and this causes a reduction in the lock-in signal. This results in the creation of a spectroscopic \enquote{saturation hole} when the probe is at $f^0_{-_i}$. Experimental results are depicted in Fig.\,\ref{fig:3holewithfit}. 
 \par
 Notice that there could be a masking effect at work in the \enquote{hole burning} experiment. When the pump is moved to a different $m_I$ transition, we also shift the saturation hole to that new $m_I$ transition (remember that $f^0_{+_i}$ and $f^0_{-_i}$ are transitions from $m_s=0$ to different $m_s$ with the same $m_I$). Consequently, an increase in the pump frequency would cause a decrease in the frequency in which the probe detects the saturation hole (e.g. when $f_+$ is moved to the right from $m_I=+1$ to $0$, the hole will move to the left from $m_I=+1$ to $0$, see Fig.\,\ref{fig:ODMRbothsideswitharrowsV2}). This imitates the effect of the dominant local magnetic fields represented by Eq.\,\eqref{eq:probevspump}. Specifically, if the pump field is on resonance with the $m_I=0$ transition from $m_s=0$ to $m_s=+1$ and we change its frequency by $+2.16\,\mathrm{MHz}$ the hole frequency will move accordingly by $-2.16\,\mathrm{MHz}$ in order to address the same $m_I$ (see Fig.\,\ref{fig:ODMRbothsideswitharrowsV2}). This masking effect is most probably what is observed in Fig.\,4(a) of Ref.\,11. Thus, care has to be taken to differentiate between the two effects, and we do this by scanning only within a specific hyperfine transition.
\par
\begin{figure}
	\centering
	\includegraphics[width=1\linewidth]{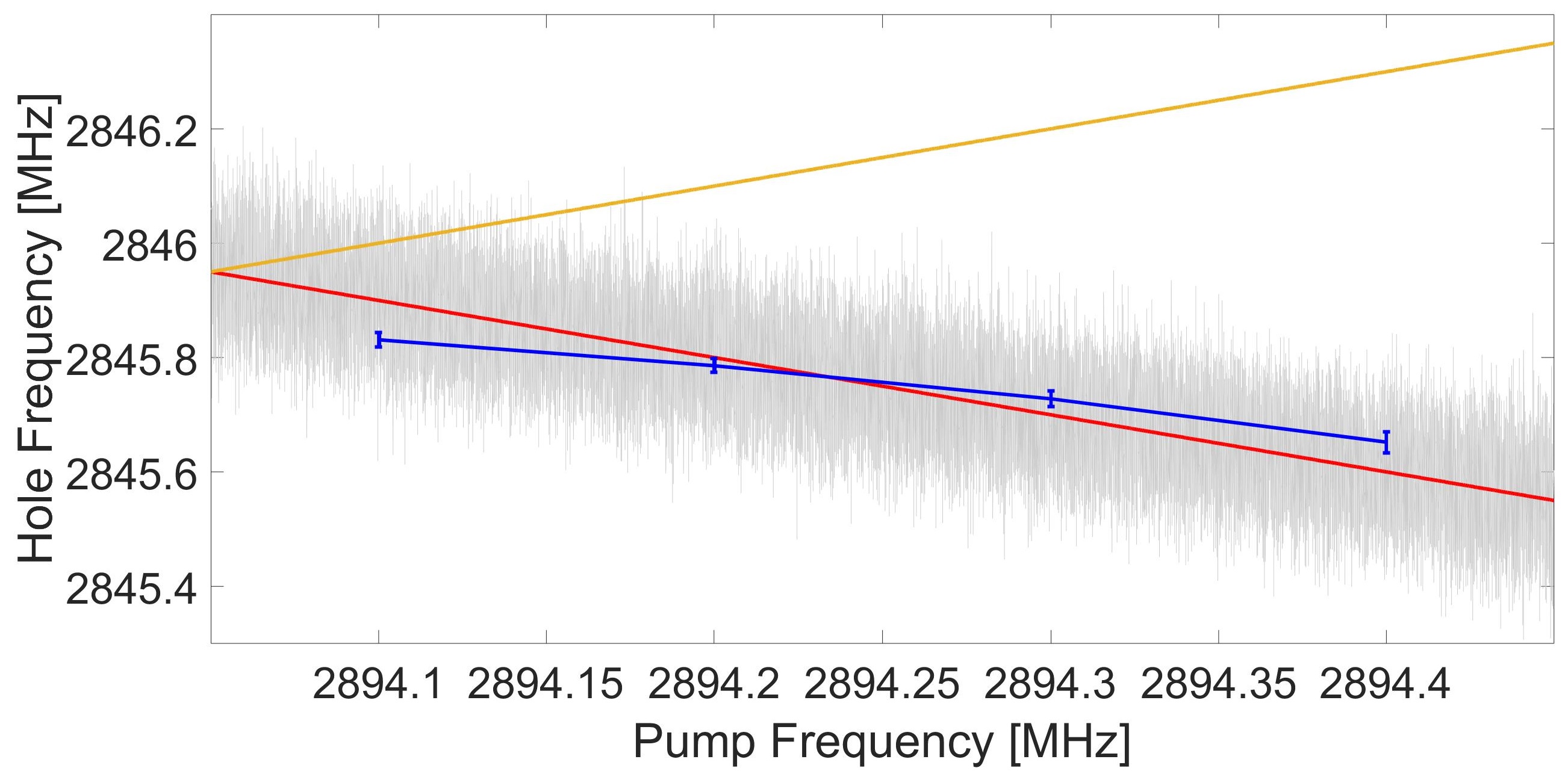}
	\caption[Migration of Hole Using the Fast Setup]{Migration of saturation hole location ($f^0_{-_0}$) as a  function of the pump ($f_+$). Data in blue, theory for dominant magnetic (electric) field in red (orange). Theory for dominant magnetic field with temperature fluctuations in gray [fluctuations do not appear for yellow line as $D_\mathrm{gs}$ does not appear in the model, i.e. Eq.\,\eqref{eq:probewithpump}]. We model the fluctuations by replacing the $D_\mathrm{gs}$ constant in Eq.\,\eqref{eq:probevspump} with a normal distribution around $2870\,\mathrm{MHz}$ having a width of $\sigma=75/2\,\mathrm{kHz}$ (equivalent to $\pm0.5\,\mathrm{K}$). The pump frequency is changed in small steps of $100\,\mathrm{kHz}$. For each pump frequency we scan the probe field to generate a spectroscopic saturation hole (see Fig.\,\ref{fig:3holewithfit}). The saturation hole lineshape is fitted and its center frequency is extracted. A linear fit returns a slope of $-0.59\pm0.21$ and $4567\pm615\,\mathrm{MHz}$ for the constant [which according to Eq.\,\eqref{eq:probevspump} is expected to be $2D_{\mathrm{gs}}=5740\,\mathrm{MHz}$]. The narrow range of data points is due to the fact that an increase in $\delta f$ with respect to $f^c_{+_0}$ decreases the signal, as we address less NVs (assuming a normal distribution of the local environment).}
	\label{fig:migration with noiseV2}
\end{figure}

In order to examine the relation between $f^0_{+_i}$ and $f^0_{-_i}$ we conduct a narrow scan of the pump frequency within the $m_I=0$ transition and measure the location of the saturation hole as a function of the pump location. For each pump frequency we scan the probe to generate and extract the saturation hole location. We vary the pump frequency in 100 kHz steps, and for each such step we repeat the probe scan and extract the hole location. The experimental results are depicted in Fig.\,\ref{fig:migration with noiseV2}. Although the results clearly favor the hypothesis of a dominant magnetic field, the value of the negative slope is found to be very sensitive to changes in the setup with slope values ranging from $-0.59$ to $-1$. Similar uncertainties were observed in the work done in Ref.\,11 \cite{private}. The negative slope in all experimental runs indicates a dominant magnetic field but due to the experimental variation, it is hard to conclude that the results are sound, and it is difficult to deduce how dominant the magnetic field effect is compared to the electric field effect. Let us briefly note that one possible reason for the inconsistency in the slope value, could be thermal fluctuations: from Eq.\,\eqref{eq:probevspump} it is evident that not only $f^0_{+_i}$ can induce a change in $f^0_{-_i}$ but so can $D_\mathrm{gs}$. $D_\mathrm{gs}$ is not constant and can have thermal fluctuations as high as $-75\,\mathrm{kHz}/\mathrm{K}$ at room temperature \cite{Acosta2010}. Thus, given that collecting statistics for each data point in Fig.\,\ref{fig:migration with noiseV2} requires a significant amount of time, even a small temperature drift may give rise to large changes, as depicted in Fig.\,\ref{fig:migration with noiseV2}. It may also be that the reason for the varying slope lies elsewhere, e.g. in some bias introduced by the measurement scheme.
\par
Whatever the reason for the unstable slope may be, it is quite evident that the magnetic field is more dominant than the electric, but also that the results are not rigorous enough.
Thus, a new and more robust method to validate Eq.\,\eqref{eq:probevspump}, by removing the dependence on $D_{\mathrm{gs}}$, is required.

\section{\enquote{Doppler-Free} Spectroscopy} \label{sec:doppler}
\begin{figure}[!ht]
	\subfloat[Pump MW is set to the $f^c_{+_{-_1}}$ resonance\label{fig:diagram_a}]{%
		\includegraphics[height=3cm,width=1\linewidth]{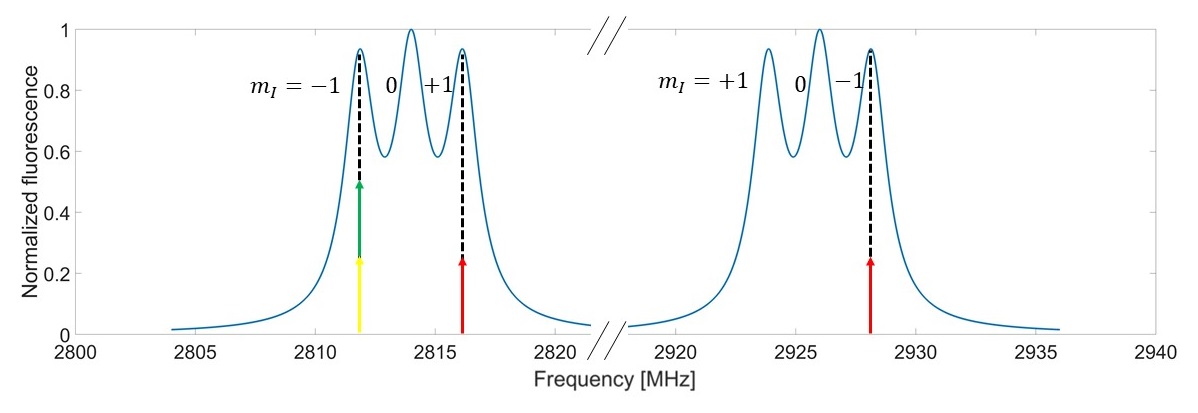}%
	}\hfill
	\subfloat[Pump MW set to be slightly detune from the $f^c_{+_0}$ resonance\label{fig:diagram_b}]{%
		\includegraphics[height=3cm,width=1\linewidth]{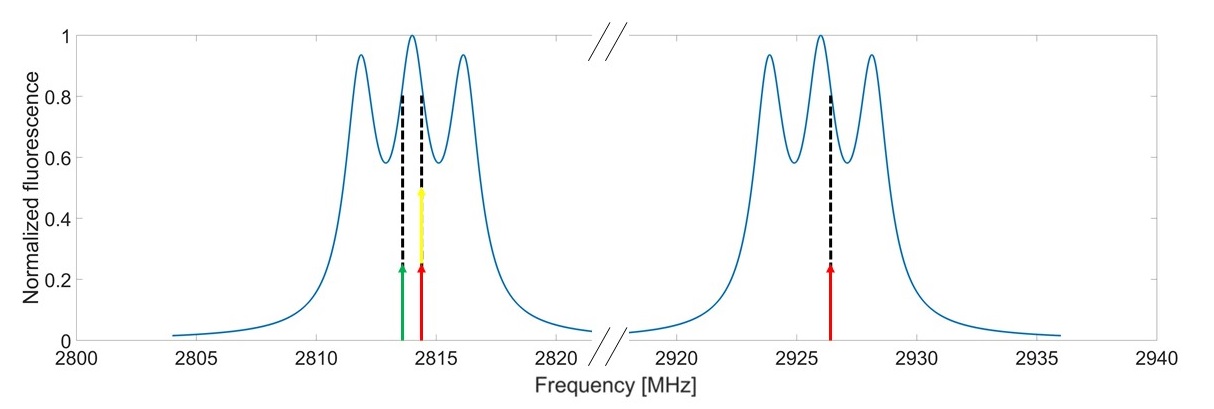}%
	}\hfill
	\subfloat[Pump MW set to the $f^c_{+_0}$ resonance\label{fig:diagram_c}]{%
		\includegraphics[height=3cm,width=1\linewidth]{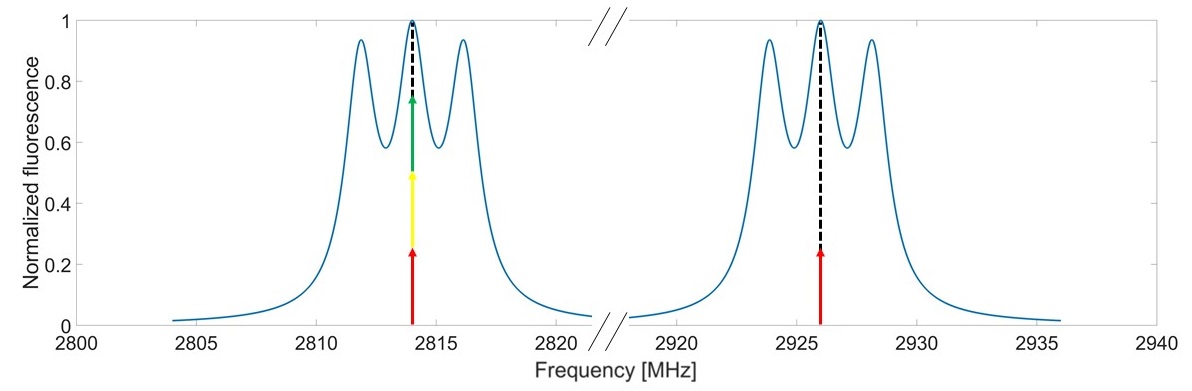}%
	}
	\caption{Schematic diagram of the different possibilities for the Doppler-free hole setup under the two hypotheses. The ODMR signal simulation of one orientation under a $20\,\mathrm{G}$ axial external field appears in blue (similar to Fig.\,\ref{fig:ODMRbothsideswitharrowsV2}), the pump field ($f_+$) is represented by the red arrow on the right, the probe field ($f_-$) is represented by the red arrow on the left, the hole location according to the hypothesis of dominant magnetic field [i.e., Eq.\,\eqref{eq:probevspump}] is represented by the green arrow, and the hole location according to the hypothesis of dominant electric field [i.e., Eq.\,\eqref{eq:probewithpump}] is represented by the yellow arrow. Dashed black lines are added for clarity. 
	(a) $f_+$ is on resonance with $f^c_{+_{-1}}$, and due to the fixed frequency gap $\Delta f$, $f_-$ is on resonance with $f^c_{-_{+1}}$,  while the hole in both hypotheses is at $f^c_{-_{-1}}$. 
	(b) For the $m_I=0$ transition, $f_+$ is shifted by $0<\delta f<1\,\mathrm{MHz}$. In the case of Eq.\,\eqref{eq:probevspump}, $f_-$ (left red arrow) does not coincide with the hole (green arrow). In contrast, in the case of Eq.\,\eqref{eq:probewithpump} (yellow arrow) the hole and the probe coincide for any value of $\delta f$ (we emphasize again the importance of $0<\delta f<1\,\mathrm{MHz}$, see text). 
	Thus, according to Eq.\,\eqref{eq:probewithpump}, as we scan the MW frequency across the entire linewidth, the hole location also varies such that it eventually covers the entire $f^0_{-0}$ transition linewidth. This results in the reduction of the ODMR signal for $m_I=0$ (Fig.\,\ref{fig:phenomenologicalfit_b}). 
	(c) $f_+$ is on resonance with $f^c_{+_0}$. Due to the fixed frequency gap $\Delta f$, $f_-$ is on resonance with $f^c_{-_0}$ and the hole in both hypotheses is at $f^c_{-_0}$. Thus, for Eq.\,\eqref{eq:probevspump} we expect to see just a narrow hole exactly at $f^c_{-0}$ (Fig.\,\ref{fig:phenomenologicalfit_a}). Finally, let us note that the transition with $m_I=0$ has higher amplitude in the diagram as we used for all three transitions a wide linewidth for clarity.} 
	\label{fig: Doppler free diagram}
\end{figure}
Our new method is analogous to Doppler-free spectroscopy in vapor in the sense that in both cases we use two radiation fields that always address different populations, except when both fields are resonant with the center of the broadened distribution. While in the vapor experiment the different populations have different velocities, here the broadening mechanism arises from the fact that each of the NVs in our ensemble can have a slightly different local environment, and consequently a different transition frequency causing inhomogeneous broadening. We replace the counter propagating fields in the vapor experiment with two MW fields, $f_+$ (pump) and $f_-$ (probe), and instead of fixing the frequencies of the radiation fields to be the same and scan the frequency as in Doppler-free spectroscopy, we rather fix the frequency gap between the two fields and then scan the frequency, so that the pump and probe move in tandem. Focusing on the $m_I=0$ transition, we fix the frequency gap to be $\Delta f=f^c_{+_0}-f^c_{-_0}$ and we scan with a CW probe while the lock-in modulated pump follows.
\par
Let us remind the reader that we denote the transition frequencies from $m_s=0$ to $m_s=\pm1$, with $m_I=i$, by $f^0_{\pm_i}$ (centered at $f^c_{\pm_i}$) and the pump (probe) field by $f_+$ ($f_-$). Due to the fixed frequency gap $\Delta f$, when the probe is on the $m_I=\pm1$ transition to $m_s=-1$, the pump is on the $m_I=\mp1$ transition to $m_s=+1$ (i.e. when $f_-=f^0_{-_{\pm1}}$ then $f_+=f^0_{+_{\mp1}}$) as depicted in Fig.\,\ref{fig:diagram_a}, and the probe has no impact as it always addresses $m_I$ states in $m_s=0$ that are different from those addressed by the modulated pump field. In such a case we expect to see a regular ODMR signal generated by the pump field as seen in Fig.\,\ref{fig:ODMRbothsideswitharrowsV2}. Notice that although $f_-$ is addressing a different $m_I$ than $f_+$, it is still on resonance, and consequently there is a reduction in the total fluorescence. However, since the lock-in detects only modulated fluorescence, this bias effect doesn't change the signal in the lock-in, and we still expect to see an ODMR signal as in Fig.\,\ref{fig:ODMRbothsideswitharrowsV2}. This holds true for both hypotheses examined in this work. The expected outcome of the two hypotheses differs only when we scan near the $m_I=0$ transition frequency.
\par
When we scan within the $m_I=0$ linewidth, both radiation fields address the same $m_I$. If $f_+ = f^c_{+_0}$ then also $f_- = f^c_{-_0}$. However, when they are not directly on resonance ($f_\pm \neq f^c_{\pm_0}$), then according to Eq.\,\eqref{eq:probevspump} they are now addressing different populations: when the probe and pump fields are \enquote{blue-detuned} with respect to the resonance transition frequency ($f_+>f^c_{+_0}$ and $f_->f^c_{-_0}$), the saturation hole is \enquote{red-detuned} due to the $-1$ ratio in Eq.\,\eqref{eq:probevspump} and the probe is not at the saturation hole frequency (see Fig.\,\ref{fig:diagram_b}), and vice versa when the pump and probe fields are red-detuned. Consequently, the probe has no impact on the lock-in signal. On the contrary, according to the hypothesis described by Eq.\,\eqref{eq:probewithpump}, if the pump and probe fields are blue (red)- detuned, so is the hole, due to the $+1$ ratio in Eq.\,\eqref{eq:probewithpump} as depicted in Fig.\,\ref{fig:diagram_b}. Thus, according to Eq.\,\eqref{eq:probewithpump} we would expect a minimum signal throughout the scan across the $m_I=0$ transition as the probe always \enquote{steals} the NV population from the pump, while in the magnetic field hypothesis [Eq.\,\eqref{eq:probevspump}] we would expect a narrow \enquote{Doppler-free} hole at $f^c_{-_0}$.
\par
This novel experimental protocol is more robust to thermal fluctuations as the latter will only induce a change in the energy difference between $m_s=0$ and $m_s=\pm1$, but the energy difference between $m_s=+1$ and $m_s=-1$ ($\Delta f$), which is a key parameter in this experiment, is not affected by it.
\par
The expected experimental lineshape of the Doppler-free hole under both hypotheses is the sum of three Lorentzians which represent the ODMR signal, and a fourth function (i.e., the hole) which is a Lorentzian with dynamic resonance frequency and amplitude (the hole frequency and amplitude changes as the pump/probe frequency changes, as will be explained later), that coincide with the transition from $m_s=0$ to $m_s=-1$ with $m_I=0$ and has a negative amplitude. Thus, the total signal has the general form
\begin{eqnarray}
S(f_-,t)=\sum_{i=1}^{3}h_i+h_4\,,
\label{eq:hole with t}
\end{eqnarray}
where 
\begin{eqnarray}
h_i(f_-)=a_i\frac{\gamma_i}{2}\frac{1}{(f_--f^c_{-_i})^2+(\frac{\gamma_i}{2})^2}
\end{eqnarray}
 represents the ODMR lineshape were $\gamma_i$ and $a_i$ are the FWHM and amplitude of the transition respectively, $f_-$ is the probe frequency and $f^c_{-_i}$ is the center of the transition frequency with $m_i=i$. The fourth function is
\begin{eqnarray}
h_4(f_-,t)=a_4\frac{\gamma_4}{2}\frac{1}{(f_--t)^2+(\frac{\gamma_4}{2})^2}\,,
\label{eq:doppler_hole_with_t}
\end{eqnarray} 
representing the hole, where $\gamma_4$ and $a_4$ are the FWHM and amplitude of the hole, respectively, and $t$ is the hole location (which changes during the scan). Based on the results of the saturation spectroscopy method, $a_4$ is expected to be negative.

\begin{figure}[!ht]
	
	\subfloat[Phenomenological simulation for dominant magnetic field ]{%	
		\label{fig:phenomenologicalfit_a}	
		\includegraphics[height=3cm,width=1\linewidth]{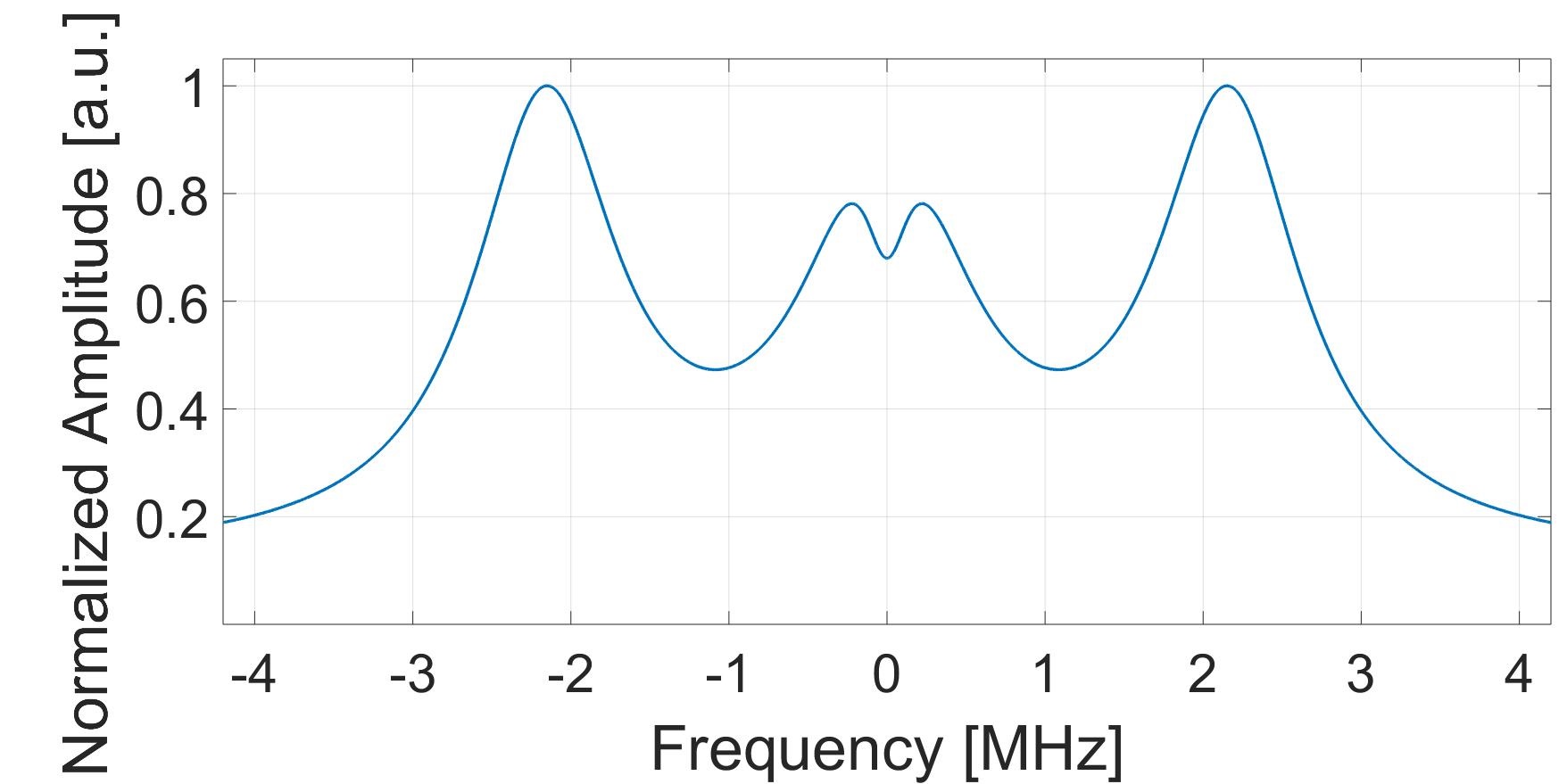}%
	}\hfill
	\subfloat[Phenomenological simulation for dominant electric field]{%	
		\label{fig:phenomenologicalfit_b}	
		\includegraphics[height=3cm,width=1\linewidth]{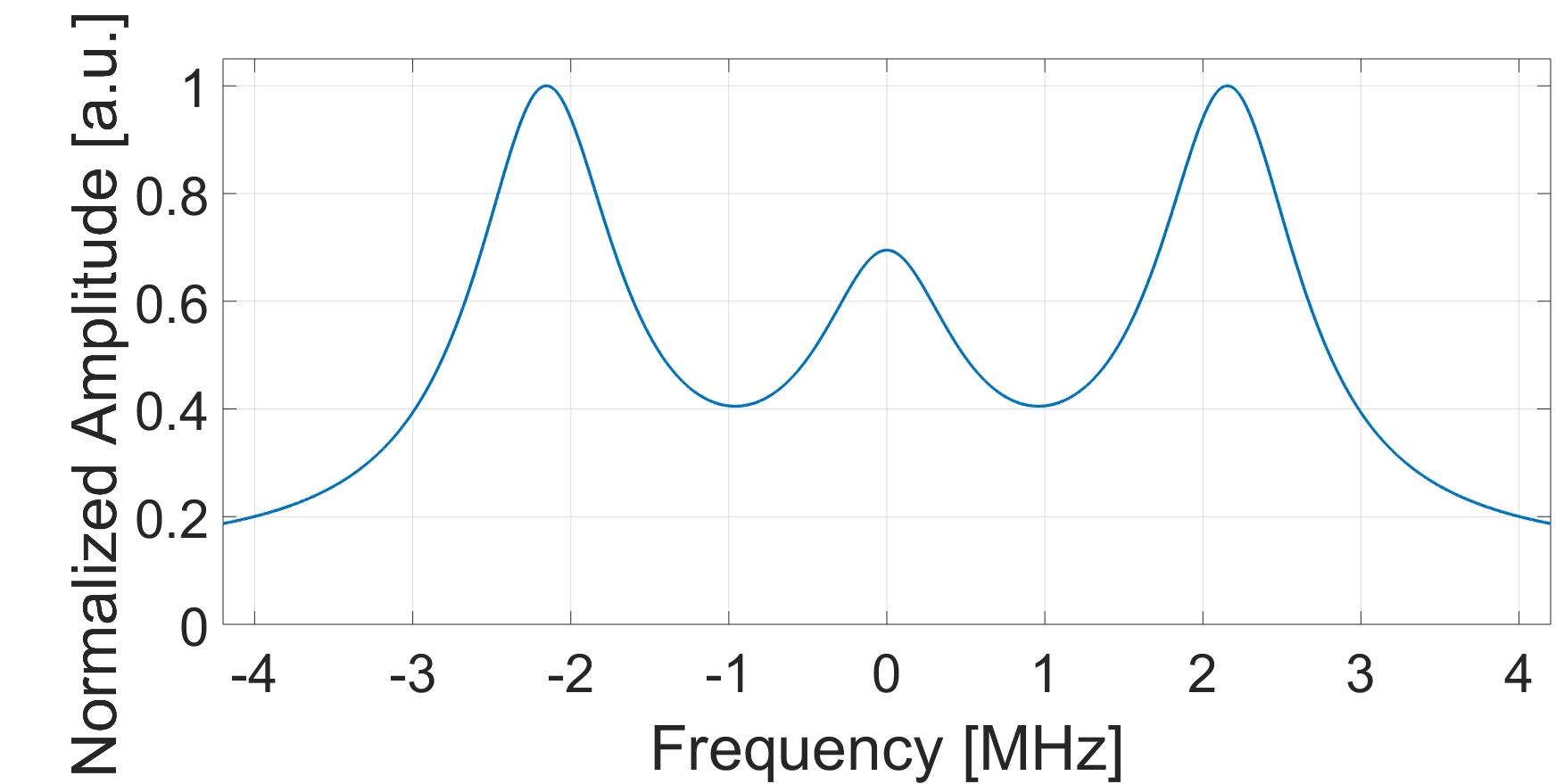}%
	}\hfill
	\subfloat[Experimental results]{%	
		\label{fig:Doppler hole}		
		\includegraphics[height=3cm,width=1\linewidth]{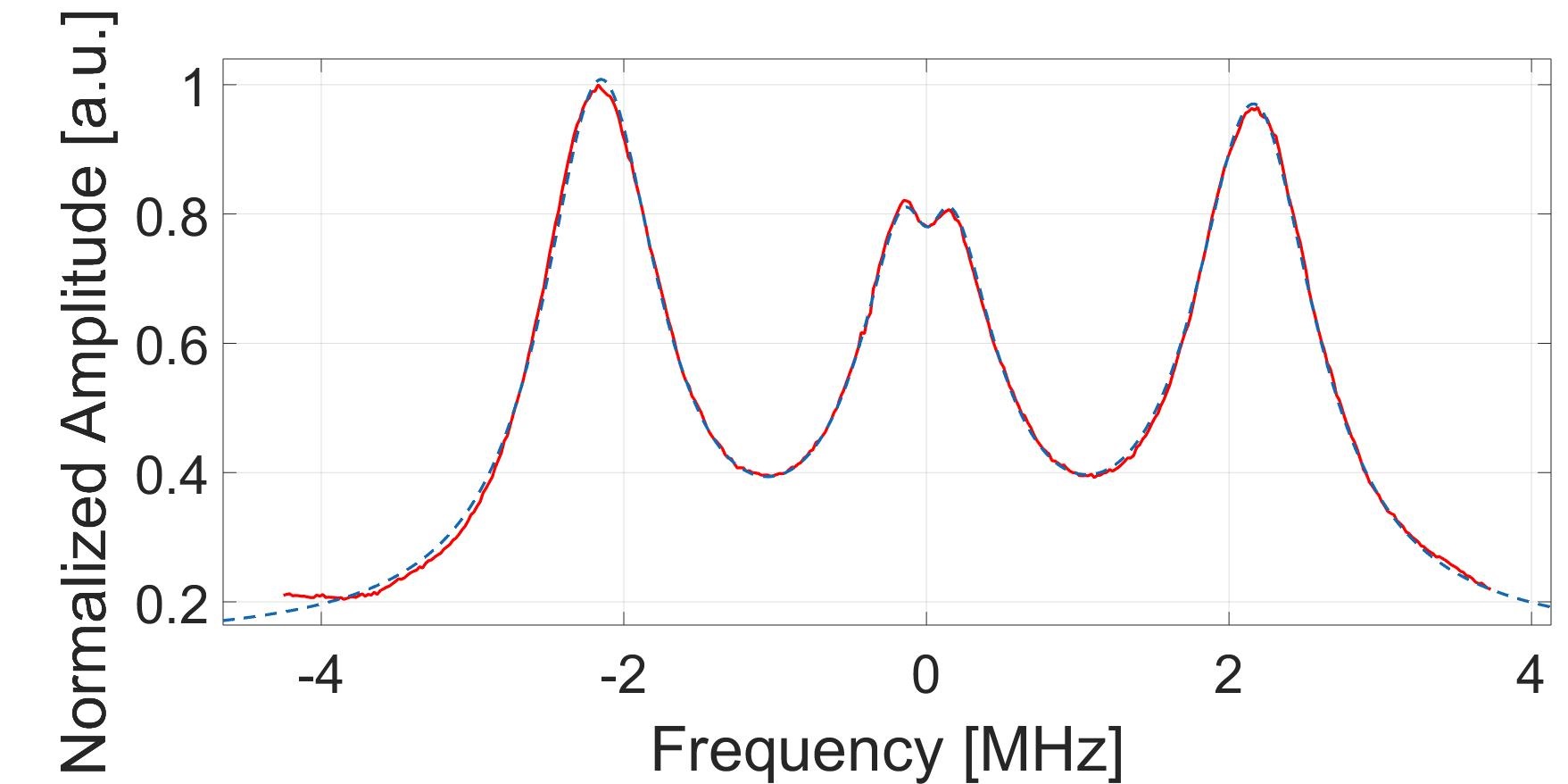}%
	}
	\caption[Doppler Type Spectroscopy]{Simulation and experimental results for Doppler-free hole spectroscopy. (a) simulation under the hypothesis of dominant magnetic field.  (b) simulation under the hypothesis of dominant electric field. (c) experimental results (modulation at $36.6\,\mathrm{kHz}$). Lock-in signal in red, fit in dashed blue. The $x$ axis represents detuning of the $f_-$ (probe) from $f^c_{-_0}$ (i.e., $\delta f$) as measured in an ODMR experiment. Experimental results are in good agreement with the dominant magnetic field simulation in Fig.\,\ref{fig:phenomenologicalfit_a} with $\alpha=-1.07\pm 0.43$. Similar results were obtained while modulating at $426\,\mathrm{Hz}$, and also for probe powers as low as $-40\,\mathrm{dbm}$. In addition, we have verified that as expected, shifting the frequency gap $\Delta f$ from the measured frequency difference between the $m_s=-1$ and $m_s=+1$ transitions moves the position of the Doppler-free hole relative to the $m_I=0$ transition center.}
\end{figure}
\par
Let us examine now the effects on $h_4$ when $f_-=f^c_{-_0}+\delta f$. The $\delta f$ shift effects the hole in two ways: a. the amplitude of the hole will decrease when the probe is detuned, since the pump is also detuned and we are now addressing less NVs which results in a reduction in the modulated signal. We may account for this effect by simply multiplying $a4$ [Eq.\,\eqref{eq:doppler_hole_with_t}] by $h2(f_-)$ [Eq.\,\eqref{eq:test}]. b. the resonance of the hole will shift by $\pm \delta f$ (Fig.\,\ref{fig:diagram_a} and Fig.\,\ref{fig:diagram_b}).
To account for the change in the hole resonance frequency during the scan we look into the tandem behavior of the MW fields. If we shift the pump frequency by $\delta f$, then the probe frequency will also move by $\delta f$ such that
$f_-=f^c_{-_0}+\delta f$.
In addition, a $\delta f$ shift in the probe frequency will cause the hole to shift by $\alpha\delta f$
\begin{eqnarray}\label{eq:hole_Doppler}
t=f^c_{-_0}+\alpha\delta f\,,
\end{eqnarray}
where $\alpha=-1$ under the hypothesis of dominant magnetic field [Eq.\,\eqref{eq:probevspump}] and $\alpha=+1$ for a dominant electric field [Eq.\,\eqref{eq:probewithpump}]
and we get
\begin{align}
t=f^c_{-_0}+\alpha(f_--f^c_{-_0})\,.
\label{eq:f_-t}
\end{align}
We can now eliminate the dependence on $t$ in Eq.\,\eqref{eq:doppler_hole_with_t} [and as a consequence, in  $S(f_-,t)$ in Eq.\,\eqref{eq:hole with t}],	as we plug Eq.\,\eqref{eq:f_-t} into Eq.\,\eqref{eq:doppler_hole_with_t} to find
\begin{eqnarray}
h_4(f_-)= \frac{a_4h_2(f_-)\frac{\gamma_4}{2}}{[f_--f^c_{-_0}-\alpha(f_--f^c_{-_0})]^2+(\frac{\gamma_4}{2})^2}\,.
\label{eq:test}
\end{eqnarray}
\par
The new parameter, $\alpha$, quantitatively distinguishes between the two hypotheses as it strongly affects the expected results.
If we set $\alpha=-1$, we can rearrange Eq.\,\eqref{eq:test} to get
\begin{eqnarray}\label{eq:mighole}
h_4(f_-,\alpha=-1)=\frac{a_5h_2(f_-)	\frac{\gamma_5}{2}}{(f_--f^c_{-_0})^2+(\frac{\gamma_5}{2})^2}\,,
\end{eqnarray}
where $a_5=0.5a_4$ and $\gamma_5=0.5\gamma_4$. The total outcome signal of such an experiment (presented in Fig.\,\ref{fig:phenomenologicalfit_a}) is thus the linear combination of $h_4(f_-,\alpha=-1)$ and an ODMR signal (Fig.\,\ref{fig:ODMRbothsideswitharrowsV2}).
\par
For a dominant electric field field [Eq.\,\eqref{eq:probewithpump}] we set $\alpha=+1$, and derive the following:
\begin{align}
h_4(f_-)=\frac{a_4h_2(f_-)}{\frac{\gamma_4}{2}}\,.
\label{eq:Doppler electric}
\end{align}
Thus, for the $m_I=0$ transition with a dominant local electric field, the Lorentzian part in Eq.\,\eqref{eq:test} is reduced to a constant. The total signal will have an ODMR-like linshape but with a reduced amplitude at the transition frequency with $m_I=0$, and furthermore the Doppler-like hole will be eliminated, since $h_2+h_4$ can be expressed as $h_2(1+\frac{a_4}{\frac{\gamma_4}{2}})$. See Fig.\,\ref{fig:phenomenologicalfit_b}.
\par
We plug the experimental values into the above models and generate a phenomenological simulation as depicted in Fig.\,\ref{fig:phenomenologicalfit_a} and Fig.\,\ref{fig:phenomenologicalfit_b}.
\par
In order to experimentally distinguish between the hypotheses, we start by calculating the frequency gap, $\Delta f$, using the results of an ODMR experiment, and fixing the frequency gap between the CW probe ($f_-$) and the modulated pump ($f_+$) to be $\Delta f$. We then scan around all hyperfine transitions with the probe (while the pump follows). All the experimental parameters are the same as in the ODMR and saturation hole experiments, except that a smaller step size for the MW fields is used. The experimental result (Fig.\,\ref{fig:Doppler hole}) shows two positive Lorentzians at the transition frequencies with $m_I=\pm1$ as both hypotheses predict, but at the transition with $m_I=0$ we observe a positive Lorentzian with a Doppler-free hole at its center. This is in good agreement with the phenomenological simulation for a dominant magnetic field environment (Fig.\,\ref{fig:phenomenologicalfit_a}).
\par
In order to also quantitatively evaluate the results, we fit the results to Eq.\,\eqref{eq:test} to find the value of $\alpha$. It is important to note that since $\alpha$ can be absorbed in $a_4$ and $\gamma_4$ (with some algebraic manipulation ) then $a_4$, $\gamma_4$ and $\alpha$ can not be determined simultaneously in the fit, and we need to predetermine the value of one of the three. We choose to set $\gamma_4=0.83\,\mathrm{MHz}$ according to the previous measurement of the hole width for the $m_I=o$ transition. 
The fit (shown in Fig.\,\ref{fig:Doppler hole}) returns a value of $\alpha=-1.07\pm 0.43$, in good agreement with the dominant magnetic field hypothesis.
\par
Combined with the results obtained using narrow saturation spectroscopy, the Doppler-free spectroscopy clearly shows that the local magnetic fields are the main cause of the inhomogeneous broadening.

\section{Summary and outlook}
\label{Summary}
We considered two possible dominant origins to inhomogeneous broadening in NV center spectroscopy: local electric fields and local magnetic fields. In order to distinguish between these two contributions, we developed a new NV spectroscopy method, which is more robust compared to hole burning spectroscopy.
\par
In the \enquote{Doppler-free} type spectroscopy we fix the frequency gap between the pump and probe MW fields to be $\Delta f$, removing the dependence on $D_{\mathrm{gs}}$, thus making the spectroscopy robust against thermal fluctuations, and moreover, enabling results to be obtained in a \enquote{single shot} instead of several independent repetitive experiments. The constant frequency gap forces a different behavior of the system depending on the origin of the inhomogeneous broadening. In the case of a dominant magnetic environment, the two MW fields interact with different NV populations, unless both fields are exactly on resonance. In this case, the scheme is expected to generate a small Doppler-free hole at the center of the transition, and this is indeed what is observed in the experiment. This proves that the origin of the inhomogeneous broadening is a variation in the local magnetic field at the vicinity of the NVs.
\\
\\
\begin{acknowledgments}
We gratefully acknowledge discussions with Pauli Kehayias, Wojciech Gawlik and Dmitry Budker.
We thank Zina Binstock for the electronics.
This work is funded in part by the Israeli Science Foundation and the German-Israeli DIP
project (Hybrid devices: FO 703/2--1) supported by the DFG.
\label{acknowledgments}
\end{acknowledgments}

%\bibliography{mybib}

\begin{thebibliography}{17}
	\expandafter\ifx\csname natexlab\endcsname\relax\def\natexlab#1{#1}\fi
	\expandafter\ifx\csname bibnamefont\endcsname\relax
	\def\bibnamefont#1{#1}\fi
	\expandafter\ifx\csname bibfnamefont\endcsname\relax
	\def\bibfnamefont#1{#1}\fi
	\expandafter\ifx\csname citenamefont\endcsname\relax
	\def\citenamefont#1{#1}\fi
	\expandafter\ifx\csname url\endcsname\relax
	\def\url#1{\texttt{#1}}\fi
	\expandafter\ifx\csname urlprefix\endcsname\relax\def\urlprefix{URL }\fi
	\providecommand{\bibinfo}[2]{#2}
	\providecommand{\eprint}[2][]{\url{#2}}
	
	\bibitem[{\citenamefont{Doherty et~al.}(2013)\citenamefont{Doherty, Manson,
			Delaney, Jelezko, Wrachtrup, and Hollenberg}}]{DOHERTY2013}
	\bibinfo{author}{\bibfnamefont{M.~W.} \bibnamefont{Doherty}},
	\bibinfo{author}{\bibfnamefont{N.~B.} \bibnamefont{Manson}},
	\bibinfo{author}{\bibfnamefont{P.}~\bibnamefont{Delaney}},
	\bibinfo{author}{\bibfnamefont{F.}~\bibnamefont{Jelezko}},
	\bibinfo{author}{\bibfnamefont{J.}~\bibnamefont{Wrachtrup}},
	\bibnamefont{and} \bibinfo{author}{\bibfnamefont{L.~C.}
		\bibnamefont{Hollenberg}}, \bibinfo{journal}{Physics Reports}
	\textbf{\bibinfo{volume}{528}}, \bibinfo{pages}{1 } (\bibinfo{year}{2013}),
	ISSN \bibinfo{issn}{0370-1573},
	\urlprefix\url{http://www.sciencedirect.com/science/article/pii/S0370157313000562}.
	
	\bibitem[{\citenamefont{Awschalom et~al.}(2013)\citenamefont{Awschalom,
			Bassett, Dzurak, Hu, and Petta}}]{Awschalom1174}
	\bibinfo{author}{\bibfnamefont{D.~D.} \bibnamefont{Awschalom}},
	\bibinfo{author}{\bibfnamefont{L.~C.} \bibnamefont{Bassett}},
	\bibinfo{author}{\bibfnamefont{A.~S.} \bibnamefont{Dzurak}},
	\bibinfo{author}{\bibfnamefont{E.~L.} \bibnamefont{Hu}}, \bibnamefont{and}
	\bibinfo{author}{\bibfnamefont{J.~R.} \bibnamefont{Petta}},
	\bibinfo{journal}{Science} \textbf{\bibinfo{volume}{339}},
	\bibinfo{pages}{1174} (\bibinfo{year}{2013}), ISSN \bibinfo{issn}{0036-8075},
	\urlprefix\url{http://science.sciencemag.org/content/339/6124/1174}.
	
	\bibitem[{\citenamefont{Hensen et~al.}(2015)\citenamefont{Hensen, Bernien,
			Dr\'eau, Reiserer, Kalb, Blok, Ruitenberg, Vermeulen, Schouten, Abell\'an
			et~al.}}]{Hensen2015}
	\bibinfo{author}{\bibfnamefont{B.}~\bibnamefont{Hensen}},
	\bibinfo{author}{\bibfnamefont{H.}~\bibnamefont{Bernien}},
	\bibinfo{author}{\bibfnamefont{A.~E.} \bibnamefont{Dr\'eau}},
	\bibinfo{author}{\bibfnamefont{A.}~\bibnamefont{Reiserer}},
	\bibinfo{author}{\bibfnamefont{N.}~\bibnamefont{Kalb}},
	\bibinfo{author}{\bibfnamefont{M.~S.} \bibnamefont{Blok}},
	\bibinfo{author}{\bibfnamefont{J.}~\bibnamefont{Ruitenberg}},
	\bibinfo{author}{\bibfnamefont{R.~F.~L.} \bibnamefont{Vermeulen}},
	\bibinfo{author}{\bibfnamefont{R.~N.} \bibnamefont{Schouten}},
	\bibinfo{author}{\bibfnamefont{C.}~\bibnamefont{Abell\'an}},
	\bibnamefont{et~al.}, \bibinfo{journal}{Nature}
	\textbf{\bibinfo{volume}{526}}, \bibinfo{pages}{682} (\bibinfo{year}{2015}),
	ISSN \bibinfo{issn}{0028-0836},
	\urlprefix\url{http://dx.doi.org/10.1038/nature15759}.
	
	\bibitem[{\citenamefont{Wolf et~al.}(2015)\citenamefont{Wolf, Neumann,
			Nakamura, Sumiya, Ohshima, Isoya, and Wrachtrup}}]{Wolf2015}
	\bibinfo{author}{\bibfnamefont{T.}~\bibnamefont{Wolf}},
	\bibinfo{author}{\bibfnamefont{P.}~\bibnamefont{Neumann}},
	\bibinfo{author}{\bibfnamefont{K.}~\bibnamefont{Nakamura}},
	\bibinfo{author}{\bibfnamefont{H.}~\bibnamefont{Sumiya}},
	\bibinfo{author}{\bibfnamefont{T.}~\bibnamefont{Ohshima}},
	\bibinfo{author}{\bibfnamefont{J.}~\bibnamefont{Isoya}}, \bibnamefont{and}
	\bibinfo{author}{\bibfnamefont{J.}~\bibnamefont{Wrachtrup}},
	\bibinfo{journal}{Phys. Rev. X} \textbf{\bibinfo{volume}{5}},
	\bibinfo{pages}{041001} (\bibinfo{year}{2015}),
	\urlprefix\url{https://link.aps.org/doi/10.1103/PhysRevX.5.041001}.
	
	\bibitem[{\citenamefont{Dolde et~al.}(2011)\citenamefont{Dolde, Fedder,
			Doherty, Nobauer, Rempp, Balasubramanian, Wolf, Reinhard, Hollenberg, Jelezko
			et~al.}}]{Dolde2011}
	\bibinfo{author}{\bibfnamefont{F.}~\bibnamefont{Dolde}},
	\bibinfo{author}{\bibfnamefont{H.}~\bibnamefont{Fedder}},
	\bibinfo{author}{\bibfnamefont{M.~W.} \bibnamefont{Doherty}},
	\bibinfo{author}{\bibfnamefont{T.}~\bibnamefont{Nobauer}},
	\bibinfo{author}{\bibfnamefont{F.}~\bibnamefont{Rempp}},
	\bibinfo{author}{\bibfnamefont{G.}~\bibnamefont{Balasubramanian}},
	\bibinfo{author}{\bibfnamefont{T.}~\bibnamefont{Wolf}},
	\bibinfo{author}{\bibfnamefont{F.}~\bibnamefont{Reinhard}},
	\bibinfo{author}{\bibfnamefont{L.~C.~L.} \bibnamefont{Hollenberg}},
	\bibinfo{author}{\bibfnamefont{F.}~\bibnamefont{Jelezko}},
	\bibnamefont{et~al.}, \bibinfo{journal}{Nat Phys}
	\textbf{\bibinfo{volume}{7}}, \bibinfo{pages}{459} (\bibinfo{year}{2011}),
	ISSN \bibinfo{issn}{1745-2473},
	\urlprefix\url{http://dx.doi.org/10.1038/nphys1969}.
	
	\bibitem[{\citenamefont{Bar-Gill et~al.}(2012)\citenamefont{Bar-Gill, Pham,
			Belthangady, Le~Sage, Cappellaro, Maze, Lukin, Yacoby, and
			Walsworth}}]{bargil2012}
	\bibinfo{author}{\bibfnamefont{N.}~\bibnamefont{Bar-Gill}},
	\bibinfo{author}{\bibfnamefont{L.}~\bibnamefont{Pham}},
	\bibinfo{author}{\bibfnamefont{C.}~\bibnamefont{Belthangady}},
	\bibinfo{author}{\bibfnamefont{D.}~\bibnamefont{Le~Sage}},
	\bibinfo{author}{\bibfnamefont{P.}~\bibnamefont{Cappellaro}},
	\bibinfo{author}{\bibfnamefont{J.}~\bibnamefont{Maze}},
	\bibinfo{author}{\bibfnamefont{M.}~\bibnamefont{Lukin}},
	\bibinfo{author}{\bibfnamefont{A.}~\bibnamefont{Yacoby}}, \bibnamefont{and}
	\bibinfo{author}{\bibfnamefont{R.}~\bibnamefont{Walsworth}},
	\bibinfo{journal}{Nature communications} \textbf{\bibinfo{volume}{3}},
	\bibinfo{pages}{858} (\bibinfo{year}{2012}).
	
	\bibitem[{\citenamefont{Takahashi et~al.}(2008)\citenamefont{Takahashi, Hanson,
			van Tol, Sherwin, and Awschalom}}]{Takahashi2008}
	\bibinfo{author}{\bibfnamefont{S.}~\bibnamefont{Takahashi}},
	\bibinfo{author}{\bibfnamefont{R.}~\bibnamefont{Hanson}},
	\bibinfo{author}{\bibfnamefont{J.}~\bibnamefont{van Tol}},
	\bibinfo{author}{\bibfnamefont{M.~S.} \bibnamefont{Sherwin}},
	\bibnamefont{and} \bibinfo{author}{\bibfnamefont{D.~D.}
		\bibnamefont{Awschalom}}, \bibinfo{journal}{Phys. Rev. Lett.}
	\textbf{\bibinfo{volume}{101}}, \bibinfo{pages}{047601}
	(\bibinfo{year}{2008}),
	\urlprefix\url{https://link.aps.org/doi/10.1103/PhysRevLett.101.047601}.
	
	\bibitem[{\citenamefont{Jensen et~al.}(2013)\citenamefont{Jensen, Acosta,
			Jarmola, and Budker}}]{Jensen2013}
	\bibinfo{author}{\bibfnamefont{K.}~\bibnamefont{Jensen}},
	\bibinfo{author}{\bibfnamefont{V.~M.} \bibnamefont{Acosta}},
	\bibinfo{author}{\bibfnamefont{A.}~\bibnamefont{Jarmola}}, \bibnamefont{and}
	\bibinfo{author}{\bibfnamefont{D.}~\bibnamefont{Budker}},
	\bibinfo{journal}{Phys. Rev. B} \textbf{\bibinfo{volume}{87}},
	\bibinfo{pages}{014115} (\bibinfo{year}{2013}),
	\urlprefix\url{https://link.aps.org/doi/10.1103/PhysRevB.87.014115}.
	
	\bibitem[{\citenamefont{Bermudez et~al.}(2011)\citenamefont{Bermudez, Jelezko,
			Plenio, and Retzker}}]{Bermudez2011}
	\bibinfo{author}{\bibfnamefont{A.}~\bibnamefont{Bermudez}},
	\bibinfo{author}{\bibfnamefont{F.}~\bibnamefont{Jelezko}},
	\bibinfo{author}{\bibfnamefont{M.~B.} \bibnamefont{Plenio}},
	\bibnamefont{and} \bibinfo{author}{\bibfnamefont{A.}~\bibnamefont{Retzker}},
	\bibinfo{journal}{Phys. Rev. Lett.} \textbf{\bibinfo{volume}{107}},
	\bibinfo{pages}{150503} (\bibinfo{year}{2011}),
	\urlprefix\url{https://link.aps.org/doi/10.1103/PhysRevLett.107.150503}.
	
	\bibitem[{\citenamefont{Mizuochi et~al.}(2009)\citenamefont{Mizuochi, Neumann,
			Rempp, Beck, Jacques, Siyushev, Nakamura, Twitchen, Watanabe, Yamasaki
			et~al.}}]{Mizuochi2009}
	\bibinfo{author}{\bibfnamefont{N.}~\bibnamefont{Mizuochi}},
	\bibinfo{author}{\bibfnamefont{P.}~\bibnamefont{Neumann}},
	\bibinfo{author}{\bibfnamefont{F.}~\bibnamefont{Rempp}},
	\bibinfo{author}{\bibfnamefont{J.}~\bibnamefont{Beck}},
	\bibinfo{author}{\bibfnamefont{V.}~\bibnamefont{Jacques}},
	\bibinfo{author}{\bibfnamefont{P.}~\bibnamefont{Siyushev}},
	\bibinfo{author}{\bibfnamefont{K.}~\bibnamefont{Nakamura}},
	\bibinfo{author}{\bibfnamefont{D.~J.} \bibnamefont{Twitchen}},
	\bibinfo{author}{\bibfnamefont{H.}~\bibnamefont{Watanabe}},
	\bibinfo{author}{\bibfnamefont{S.}~\bibnamefont{Yamasaki}},
	\bibnamefont{et~al.}, \bibinfo{journal}{Phys. Rev. B}
	\textbf{\bibinfo{volume}{80}}, \bibinfo{pages}{041201}
	(\bibinfo{year}{2009}),
	\urlprefix\url{https://link.aps.org/doi/10.1103/PhysRevB.80.041201}.
	
	\bibitem[{\citenamefont{Kehayias et~al.}(2014)\citenamefont{Kehayias, Mr\'ozek,
			Acosta, Jarmola, Rudnicki, Folman, Gawlik, and Budker}}]{pauli2014}
	\bibinfo{author}{\bibfnamefont{P.}~\bibnamefont{Kehayias}},
	\bibinfo{author}{\bibfnamefont{M.}~\bibnamefont{Mr\'ozek}},
	\bibinfo{author}{\bibfnamefont{V.~M.} \bibnamefont{Acosta}},
	\bibinfo{author}{\bibfnamefont{A.}~\bibnamefont{Jarmola}},
	\bibinfo{author}{\bibfnamefont{D.~S.} \bibnamefont{Rudnicki}},
	\bibinfo{author}{\bibfnamefont{R.}~\bibnamefont{Folman}},
	\bibinfo{author}{\bibfnamefont{W.}~\bibnamefont{Gawlik}}, \bibnamefont{and}
	\bibinfo{author}{\bibfnamefont{D.}~\bibnamefont{Budker}},
	\bibinfo{journal}{Phys. Rev. B} \textbf{\bibinfo{volume}{89}},
	\bibinfo{pages}{245202} (\bibinfo{year}{2014}),
	\urlprefix\url{https://link.aps.org/doi/10.1103/PhysRevB.89.245202}.
	
	\bibitem[{\citenamefont{Manson et~al.}(2006)\citenamefont{Manson, Harrison, and
			Sellars}}]{Manson2006}
	\bibinfo{author}{\bibfnamefont{N.~B.} \bibnamefont{Manson}},
	\bibinfo{author}{\bibfnamefont{J.~P.} \bibnamefont{Harrison}},
	\bibnamefont{and} \bibinfo{author}{\bibfnamefont{M.~J.}
		\bibnamefont{Sellars}}, \bibinfo{journal}{Phys. Rev. B}
	\textbf{\bibinfo{volume}{74}}, \bibinfo{pages}{104303}
	(\bibinfo{year}{2006}),
	\urlprefix\url{https://link.aps.org/doi/10.1103/PhysRevB.74.104303}.
	
	\bibitem[{\citenamefont{Goldman et~al.}(2015)\citenamefont{Goldman, Sipahigil,
			Doherty, Yao, Bennett, Markham, Twitchen, Manson, Kubanek, and
			Lukin}}]{Goldman2015}
	\bibinfo{author}{\bibfnamefont{M.~L.} \bibnamefont{Goldman}},
	\bibinfo{author}{\bibfnamefont{A.}~\bibnamefont{Sipahigil}},
	\bibinfo{author}{\bibfnamefont{M.~W.} \bibnamefont{Doherty}},
	\bibinfo{author}{\bibfnamefont{N.~Y.} \bibnamefont{Yao}},
	\bibinfo{author}{\bibfnamefont{S.~D.} \bibnamefont{Bennett}},
	\bibinfo{author}{\bibfnamefont{M.}~\bibnamefont{Markham}},
	\bibinfo{author}{\bibfnamefont{D.~J.} \bibnamefont{Twitchen}},
	\bibinfo{author}{\bibfnamefont{N.~B.} \bibnamefont{Manson}},
	\bibinfo{author}{\bibfnamefont{A.}~\bibnamefont{Kubanek}}, \bibnamefont{and}
	\bibinfo{author}{\bibfnamefont{M.~D.} \bibnamefont{Lukin}},
	\bibinfo{journal}{Phys. Rev. Lett.} \textbf{\bibinfo{volume}{114}},
	\bibinfo{pages}{145502} (\bibinfo{year}{2015}),
	\urlprefix\url{https://link.aps.org/doi/10.1103/PhysRevLett.114.145502}.
	
	\bibitem[{\citenamefont{Robledo et~al.}(2011)\citenamefont{Robledo, Bernien,
			van~der Sar, and Hanson}}]{Robledo2011}
	\bibinfo{author}{\bibfnamefont{L.}~\bibnamefont{Robledo}},
	\bibinfo{author}{\bibfnamefont{H.}~\bibnamefont{Bernien}},
	\bibinfo{author}{\bibfnamefont{T.}~\bibnamefont{van~der Sar}},
	\bibnamefont{and} \bibinfo{author}{\bibfnamefont{R.}~\bibnamefont{Hanson}},
	\bibinfo{journal}{New Journal of Physics} \textbf{\bibinfo{volume}{13}},
	\bibinfo{pages}{025013} (\bibinfo{year}{2011}),
	\urlprefix\url{http://stacks.iop.org/1367-2630/13/i=2/a=025013}.
	
	\bibitem[{\citenamefont{Mrozek et~al.}(2016)\citenamefont{Mrozek,
			Wojciechowski, Rudnicki, Zachorowski, Kehayias, Budker, and
			Gawlik}}]{Gawlik2016}
	\bibinfo{author}{\bibfnamefont{M.}~\bibnamefont{Mrozek}},
	\bibinfo{author}{\bibfnamefont{A.~M.} \bibnamefont{Wojciechowski}},
	\bibinfo{author}{\bibfnamefont{D.~S.} \bibnamefont{Rudnicki}},
	\bibinfo{author}{\bibfnamefont{J.}~\bibnamefont{Zachorowski}},
	\bibinfo{author}{\bibfnamefont{P.}~\bibnamefont{Kehayias}},
	\bibinfo{author}{\bibfnamefont{D.}~\bibnamefont{Budker}}, \bibnamefont{and}
	\bibinfo{author}{\bibfnamefont{W.}~\bibnamefont{Gawlik}},
	\bibinfo{journal}{Phys. Rev. B} \textbf{\bibinfo{volume}{94}},
	\bibinfo{pages}{035204} (\bibinfo{year}{2016}),
	\urlprefix\url{https://link.aps.org/doi/10.1103/PhysRevB.94.035204}.
	
	\bibitem[{\citenamefont{{Pauli Kehayias}}(private comminication)}]{private}
	\bibinfo{author}{\bibnamefont{{Pauli Kehayias}}} (\bibinfo{year}{private
		comminication}).
	
	\bibitem[{\citenamefont{Acosta et~al.}(2010)\citenamefont{Acosta, Bauch,
			Ledbetter, Waxman, Bouchard, and Budker}}]{Acosta2010}
	\bibinfo{author}{\bibfnamefont{V.~M.} \bibnamefont{Acosta}},
	\bibinfo{author}{\bibfnamefont{E.}~\bibnamefont{Bauch}},
	\bibinfo{author}{\bibfnamefont{M.~P.} \bibnamefont{Ledbetter}},
	\bibinfo{author}{\bibfnamefont{A.}~\bibnamefont{Waxman}},
	\bibinfo{author}{\bibfnamefont{L.-S.} \bibnamefont{Bouchard}},
	\bibnamefont{and} \bibinfo{author}{\bibfnamefont{D.}~\bibnamefont{Budker}},
	\bibinfo{journal}{Phys. Rev. Lett.} \textbf{\bibinfo{volume}{104}},
	\bibinfo{pages}{070801} (\bibinfo{year}{2010}),
	\urlprefix\url{https://link.aps.org/doi/10.1103/PhysRevLett.104.070801}.
	
\end{thebibliography}

\end{document}